\documentclass[letterpaper]{article} 
\usepackage{aaai25}  
\nocopyright
\usepackage{times}  
\usepackage{helvet}  
\usepackage{courier}  
\usepackage[hyphens]{url}  
\usepackage{graphicx} 
\usepackage{natbib}  
\usepackage{caption} 
\usepackage{algorithm}
\usepackage{algorithmic}
\usepackage{amssymb}
\usepackage{amsmath}
\usepackage{multirow}

\usepackage{newfloat}
\usepackage{listings}
\usepackage{booktabs}
\DeclareCaptionStyle{ruled}{labelfont=normalfont,labelsep=colon,strut=off} 
\floatstyle{ruled}
\newfloat{listing}{tb}{lst}{}
\floatname{listing}{Listing}
\title{FreqINR: Frequency Consistency for Implicit Neural Representation with Adaptive DCT Frequency Loss}
\author{
    Meiyi Wei\textsuperscript{\rm 1}
     Liu Xie,\textsuperscript{\rm 1}
     Ying Sun,\textsuperscript{\rm 1}
    Gang Chen\equalcontrib\textsuperscript{\rm 1}
}
\affiliations{
    \textsuperscript{\rm 1}School of Cyber Science and Engineering, Wuhan University\\    

}

\usepackage{bibentry}

\begin{document}
\frenchspacing
\maketitle

\begin{abstract}
Recent works on local Implicit Neural Representation (INR) have highlighted its exceptional performance in depicting images across arbitrary resolutions. However, frequency divergence between high-resolution (HR) and ground-truth images persists at various scales, particularly at larger scales, leading to severe artifacts and blurring in HR image.
In this paper, we propose Frequency Consistency for Implicit Neural Representation (FreqINR), an innovative Arbitrary-scale Super-resolution method designed to refine detailed textures by ensuring spectral consistency throughout training and inference. During training, we introduce Adaptive Discrete Cosine Transform Frequency Loss (ADFL) to narrow the frequency gap between HR and ground-truth images, employing 2-Dimensional DCT bases and dynamically focusing on challenging frequencies. During inference, we extend the receptive field to maintain spectral coherence between low-resolution (LR) and ground-truth images, which is essential for model to ”hallucinate“ high-frequency details from LR counterpart.
Experimental results demonstrate that, as a lightweight operation, FreqINR achieves state-of-the-art (SOTA) performance compared to existing Arbitrary-scale SR methods. Additionally, our implementation shows significant advantages in terms of computational efficiency. Our code will be made publicly available.
\end{abstract}

%
\section{Introduction}

\begin{figure}[t]
\centering
\includegraphics[width=\linewidth]{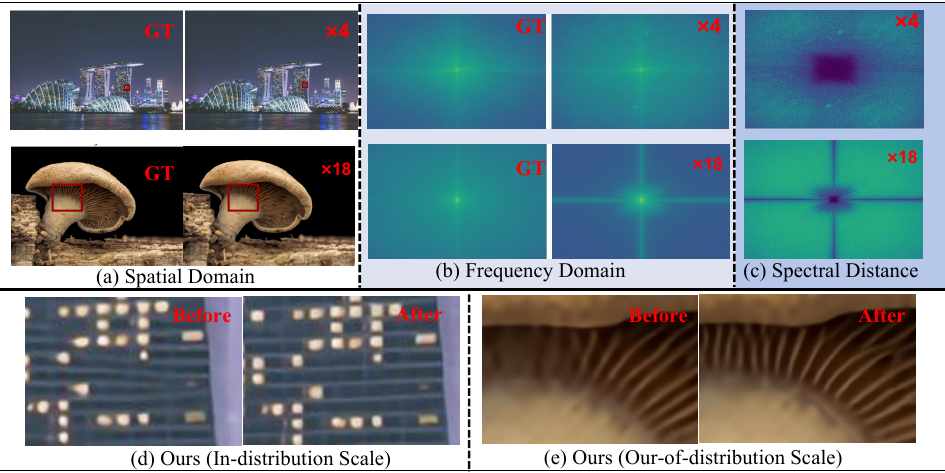} 
\caption{Main concept of FreqINR. In (b) and (c), bright areas represent strong frequencies, while dark areas indicate weak ones. (a) shows ground-truth and HR images generated by EDSR-baseline-LIIF \cite{Chen_2021_CVPR} before applying FreqINR, with $\times 4$ (in-distribution) in the first row and $\times 18$ (out-of-distribution) in the second row. (b) illustrates the frequency domain transformation. (c) presents the frequency distances. (d) and (e) display the visual results of our FreqINR at $\times 4$ and $\times 18$ scales, respectively.}
\label{Fig:fig0}
\end{figure}
\begin{figure*}[t]
\centering
\includegraphics[width=\textwidth]{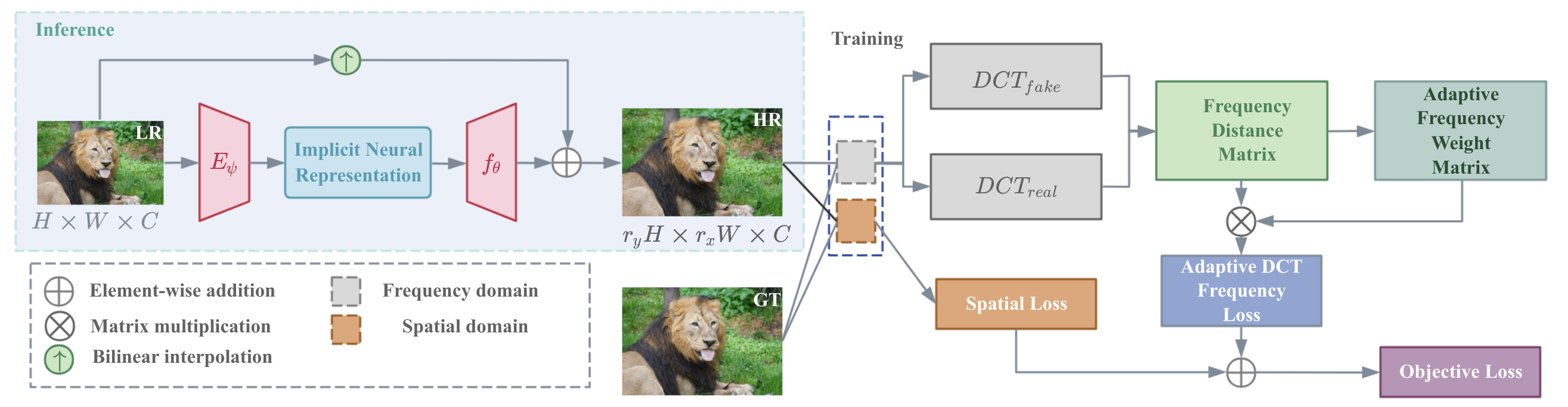} 
\caption{Overview of FreqINR. The inference process for INR-based models (light blue) is guided by our core component, \textit{Adaptive DCT Frequency Loss} (dark blue), which leverages the \textit{Frequency Distance Matrix} (light green) and the \textit{Adaptive Frequency Weight Matrix} (dark green) to dynamically enhance fine detail learning.}
\label{Fig:fig1}
\end{figure*}

Single image super-resolution (SISR) aims to restore high-resolution (HR) images from low-resolution (LR) counterparts. Traditional deep learning methods using direct upsampling in end-to-end architectures perform well for fixed upsampling scales \cite{chen2021pre, Liang_2021_ICCV, Zhou_2023_ICCV, Lim_2017_CVPR_Workshops, Liu_2021_ICCV, Magid_2021_ICCV, Shi_2016_CVPR, Zhang_2018_ECCV, Zhang_2018_CVPR}. However, these methods require retraining for each scale factor, which is impractical in real-world scenarios.

Arbitrary-scale super-resolution (ASSR) resolve this by providing multi-scale compatibility within a single network \cite{Hu_2019_CVPR, Wang_2021_ICCV}. Recent advancements in Implicit Neural Representation (INR) have shown promise by representing images as continuous functions rather than discrete grids \cite{Chen_2021_CVPR, Lee_2022_CVPR, Wei_2023_CVPR, Chen_2023_CVPR}. Techniques like LIIF \cite{Chen_2021_CVPR} and LTE \cite{Lee_2022_CVPR} have tackled spectral bias but still face performance limitations with multi-layer perceptrons (MLPs). Meanwhile, SRNO \cite{Wei_2023_CVPR} and CLIT \cite{Chen_2023_CVPR} made use of attention mechanisms to capture non-local correlations. In a nutshell, these Implicit Neural Representation (INR)-based methods assume that latent variables are evenly distributed in the spatial domain, implying that images at arbitrary resolutions share similar textures within local region. Therefore, achieving frequency domain consistency between the given LR and the ground-truth images enhances produced images. Even so, as shown in Fig.~\ref{Fig:fig0}, a frequency gap remains, causing deformation and blurring, particularly at larger scales. These frequency discrepancies hinder accurate restoration of genuine textures.

Current research \cite{Jiang_2021_ICCV, Kim_2023_WACV} in image reconstruction shows that incorporating frequency loss to minimize spectral discrepancies helps the model capture delicate textures. In contrast to fixed-scale image reconstruction, Fig.~\ref{Fig:fig0} illustrates that spectral discrepancies become more pronounced at larger SR scales, while they are less significant at smaller scales and are mainly concentrated in high frequencies. Generally, in Arbitrary-Scale SR, although frequency domain features are consistent across scales, the model needs to focus on different frequency distributions and ranges between HR and gourd-truth images for varying scales. To address this, we aim to direct the INR-based models to dynamically concentrate on challenging frequencies, especially high-frequency components.

Building on these observations, we introduce Frequency Consistency for Implicit Neural Representation (FreqINR), a novel method designed to enhance image quality in arbitrary-scale SR. FreqINR significantly improves advanced INR-based models by adaptively addressing frequency divergence and restoring detailed high-frequency patterns during both training and inference.
During training, FreqINR integrates frequency consistency and global spatial correlations between HR and ground-truth images into the Implicit Neural Function framework through a unified objective function that combines pixel-wise loss with our Adaptive Discrete Cosine Transform Frequency Loss (ADFL). Specifically, we use Discrete Cosine Transform (DCT) to convert spatial data into the frequency domain, employ a Frequency Distance Matrix (FDM) to manage varying frequencies, and apply an Adaptive Frequency Weighting Matrix (AFWM) to dynamically adjust weights based on amplitude information.
During inference, to match the frequency distribution of input LR with ground-truth images, we extend the encoder’s receptive field without bringing computational overhead. Experiments verify that FreqINR markedly improves image quality by enhancing frequency alignment and spatial coherence. In summary, our main contributions are as follows:
\begin{itemize}
\item We propose FreqINR, a generic framework that ameliorates the quality of target HR images by maintaining frequency consistency across all scale factors during both training and inference.
\item We introduce an Adaptive Discrete Cosine Transform Frequency Loss (ADFL) for training, which adaptively narrows the frequency gap of HR and ground-truth images, and alleviates function overshooting caused by using Fourier basis functions for image transformation.
\item We expand the receptive field of encoders during inference to ensure frequency coherence between LR and ground-truth images, improving latent feature extraction without added computational cost.
\item Experiments points out any INR-based backbone leveraged by FreqINR deliver genuine high-frequency details and boost performance in arbitrary-scale SR.
\end{itemize}

\section{Related Works}
This section reviews advancements in super-resolution (SR), discusses local implicit neural representations (INR) for arbitrary-scale SR, and analyzes the impact of spectral bias in INR. For a better understanding, we include a preliminary review of Implicit Neural Representations and Focal Frequency Loss in the supplementary material.
\subsubsection{Deep SR architecture.}

Remarkable progress have been made in Single Image Super-Resolution (SISR) despite ongoing challenges \cite{chen2021pre, Liang_2021_ICCV, Zhou_2023_ICCV, Lim_2017_CVPR_Workshops, Liu_2021_ICCV, Magid_2021_ICCV, Shi_2016_CVPR, Zhang_2018_ECCV, Zhang_2018_CVPR}. Early efforts primarily focused on enhancing upsampling through CNN-based models \cite{Lim_2017_CVPR_Workshops, Shi_2016_CVPR, Zhang_2018_ECCV, Zhang_2018_CVPR}. Recently, the success of self-attention mechanisms \cite{Liu_2021_ICCV, vit} has led to the adoption of Transformers \cite{chen2021pre, Liang_2021_ICCV, Zhou_2023_ICCV, Magid_2021_ICCV} in SR frameworks, such as SwinIR \cite{Liang_2021_ICCV} and SRFormer \cite{Zhou_2023_ICCV}, for high-level feature extraction. A notable limitation of these approaches is their reliance on fixed scaling factors, necessitating separate models for each, which is inefficient. To overcome this, methods like Meta-SR \cite{Hu_2019_CVPR}, ArbSR \cite{Wang_2021_ICCV}, and SRWarp \cite{Son_2021_CVPR} have been developed to create unified models for arbitrary-scale SR. However, their effectiveness diminishes for scaling factors outside the distribution of the training data. 

\subsubsection{Local Implicit Neural Representation.}

Inspired by implicit neural function, representing images as RGB-valued functions and utilizing a shared local Implicit Neural Representation (INR) are shined in handling arbitrary-scale SR \cite{Chen_2021_CVPR,Lee_2022_CVPR,Wei_2023_CVPR,Chen_2023_CVPR, ultrasr, 30}. LIIF \cite{Chen_2021_CVPR} first adopts a distinctive strategy by a Multilayer Perceptron (MLP) as a local implicit function. It predicts an RGB value for a given coordinate by leveraging nearby LR features and a cell size. UltraSR \cite{ultrasr} and IPE \cite{30} build on this by using embeddings instead of coordinates, addressing the spectral bias inherent in MLPs. Further advancing this paradigm, LTE \cite{Lee_2022_CVPR} refines this approach with a local texture estimator that translates coordinates into the Fourier domain. Despite these advances, decoders based on MLPs often struggle to accurately represent arbitrary images. Recent methods such as SRNO \cite{Wei_2023_CVPR} and CLIT \cite{Chen_2023_CVPR} incorporate attention mechanisms to capture non-local features from low-resolution images but still rely on coordinate-based embeddings \cite{Gu_2021_CVPR}, which lacks consideration of important global frequency information. Yet, these methods rely only on per-pixel L1 loss \cite{32, Isola_2017_CVPR}, which weights all pixels equally and favors low-frequency over high-frequency components. To address this, we leverage frequency domain consistency over space of INR and introduce a dynamic frequency loss to balance the frequency components.

\subsubsection{Spectral bias.}

Frequency analysis uncovers a phenomenon known as spectral bias \cite{NEURIPS2020_55053683, 32, 33}, where neural networks tend to favor learning low-frequency functions. Meanwhile, we have mentioned diverse applications techniques aim to recover missing high frequencies in ASSR \cite{Lee_2022_CVPR, Wei_2023_CVPR, Chen_2023_CVPR}. Nonetheless, unlike uniform weighting pixel in the spatial domain, frequency coordinates are related to all spatial pixels, above methods disregard independently address the spectrum distribution. Simultaneously, frequency domain analysis has proven vital for restoring fine image details in image reconstruction \cite{Jiang_2021_ICCV, Kim_2023_WACV}. FFL \cite{Jiang_2021_ICCV} rationally proves that reducing the frequency domain gap can notable ameliorate image synthesis quality, and it reveals that mining hard samples for difficult frequencies can address spectrum biases. Building on this, wavelet transforms were utilized \cite{Kim_2023_WACV}, incorporating a module for multi-bandwidth analysis. Even so, distinct from fixed-scale image reconstruction, frequency distribution becomes increasingly imbalanced across different scales. To solve this issue, we propose a loss function that adapts to these frequency variations in arbitrary-scale SR.


\section{Method}
In this section, we describe the key techniques of FreqINR: Adaptive DCT Frequency Loss (ADFL) for training and Enhanced Receptive Field Encoder for inference. The overall architecture of FreqINR is illustrated in Fig.~\ref{Fig:fig1}. 

\subsection{Adaptive DCT Frequency Loss}
During training, we introduce Adaptive DCT Frequency Loss (ADFL). First, we represent image by DCT bases. Then, we employ the Frequency Distance Matrix (FDM) to guide the Adaptive Frequency Weighting Matrix (AFWM) in dynamically minimizing spectral discrepancies of generated HR and ground-truth. Finally, we integrate ADFL into the per-pixel spatial loss to form our final objective function.
\subsubsection{Frequency Representation of Images.}

Motivated by the FFL \cite{Jiang_2021_ICCV}, which highlights how frequency distribution affects spatial image quality, we find that minimizing the spectral distance enables generative models to produce images that closely resemble the ground-truth. Therefore, transforming continuous spatial RGB signals into discrete signals and reducing the frequency gaps is a promising approach to enhancing image quality of generated HR. 

For representing image textures with frequency basis functions across arbitrary resolutions, LTE \cite{Lee_2022_CVPR} addresses the issue of spectral bias by using a texture estimator. Nonetheless, representing continuous signals with a finite sum of Fourier basis functions can lead to overshoots at discontinuities and step functions, a problem known as the Gibbs phenomenon or ringing artifacts. To overcome these issues, especially at out-of-distribution scales, we use the Discrete Cosine Transform (DCT) \cite{1672377} instead of the proposed loss in \cite{Jiang_2021_ICCV}. The DCT not only smooths the algorithm to mitigate such effects but also provides a more energy-concentrated representation, which facilitates better artifact removal by discarding redundant or less important information that may not be perceptible to the human eye. Specifically, we apply the 2D DCT to obtain frequency domain representation of given image:

\begin{equation}
\fontsize{9pt}{\baselineskip}\selectfont
\begin{split}
   F(u, v) &= C(u)C(v) \sqrt{\frac{2}{MN}} \sum_{x=0}^{M-1}\sum_{y=0}^{N-1}f(x, y) \\
   &\quad \times \cos\left[\frac{\pi}{M}u\left(x+\frac{1}{2}\right)\right] \cos\left[\frac{\pi}{N}v\left(y+\frac{1}{2}\right)\right].
\end{split}
   \label{eq:Eq0}
\end{equation}

where the size of the image is $M\times N$. $(x, y)$ indicates the image coordinates of spatial domain. $f(x,y)$ is the RGB value. $(u,v)$ denotes the coordinates of a spatial frequency on the frequency spectrum. $F(u, v)$ is the transformed real frequency value. Compared to DFT, DCT only has one low-frequency point, which occurs when $k=0$. When performing DCT transformations, in order to ensure the orthogonality of the transform bases, a constant $C(k)$ is introduce:

 \begin{equation}
   C(k) = \begin{cases} \frac{1}{\sqrt{2}}, & \text{$k=0$}, \\  1, & \text{$1\le k \le N-1$} \end{cases}, \ where \ k=u, v
      \label{eq:Eq1}
\end{equation}        

\subsubsection{Frequency Distance Matrix.}

In FFL \cite{Jiang_2021_ICCV}, frequency distance measures image differences, but DCT's magnitude spectrum differs from DFT's. The spectrum weight in \cite{Jiang_2021_ICCV} ignores the frequency values' dynamic range, where high frequencies are much smaller than low ones. To improve this, we adjust the frequency weighting function using the absolute value of the logarithm of the spectrum distance. This adjustment increases the weight for lower scale distances. The weight is defined as:

 \begin{equation}
 \fontsize{9pt}{\baselineskip}\selectfont
   FDM(u, v) =w_0(u, v) = \vert \log{(\vert F_r(u, v)-F_f(u, v)\vert)}\vert ^\alpha.
    \label{eq:Eq2}
\end{equation}      

The frequency distance between reference points in image pairs \((F_r, F_f)\) is defined in Euclidean space at spectral position \((u, v)\) as the frequency distance matrix. Here, \(\alpha\) serves as a control factor to adjust changes in each section. Similar to FFL, we normalize the weight matrix by dividing \(w_0(u, v)\) by its maximum value \(\max_{w_0}\), and refer to it as \(w_n(u, v)\) from now on.

\begin{figure}[t]
\centering
\includegraphics[width=\columnwidth]{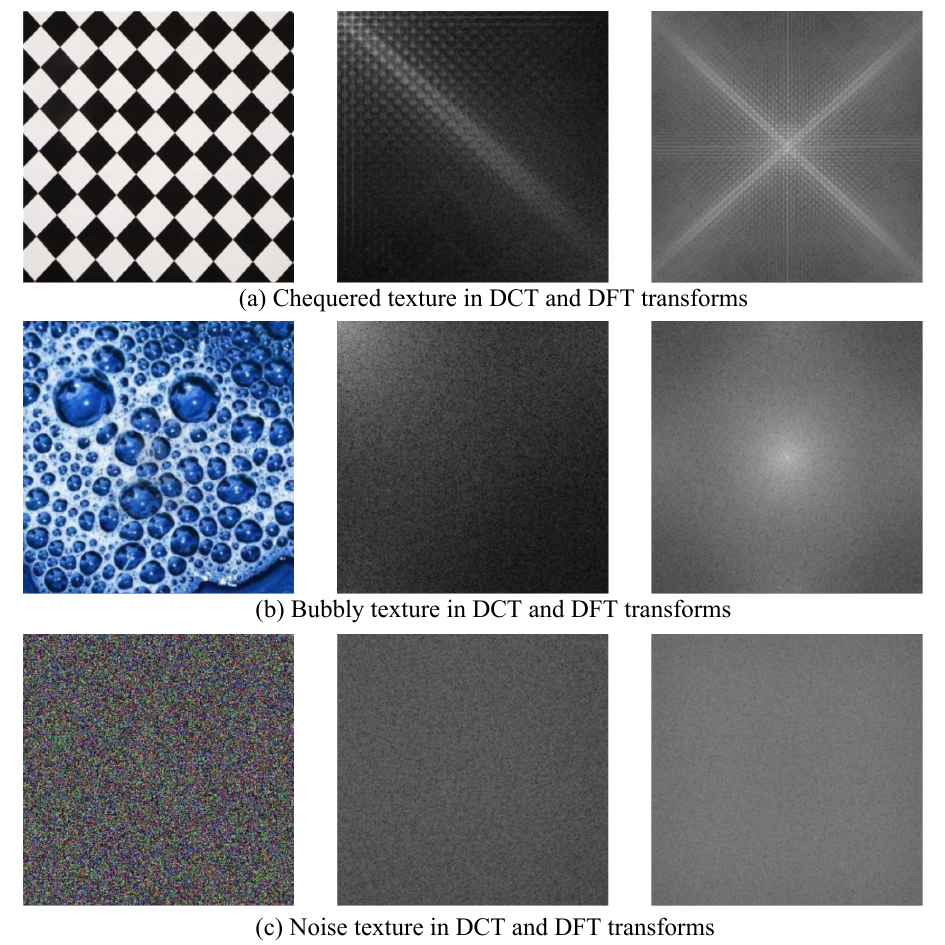} 
\caption{Comparison of DCT and DFT distributions. Unlike DFT, DCT places low frequencies in the upper-left corner. (a) The \textit{Chequered} texture primarily consists of horizontal and vertical details. (b) The \textit{Bubbly} texture includes both fine and coarse details, reflecting the frequency distribution of natural images. (c) The \textit{Noise} texture is common and typically needs removal.}
\label{Fig:fig2}
\end{figure}
\subsubsection{Adaptive Frequency Weighting Matrix.}

As shown in Fig.~\ref{Fig:fig2}, the distribution of DCT is difference from DFT, being more energy-concentrated with a single low-frequency point at the upper left corner. For effective image reconstruction, emphasizing high-frequency details is crucial. To address this, we introduce a control coefficient, to enforce the model in focusing on high-frequency components. Thus, we use a weight mask \(M_{DCTL}\) that aligns with spectrum distribution of DCT, defined as follows:
\begin{equation}
\fontsize{9pt}{\baselineskip}\selectfont
   M_{DCTL}(u, v) = \begin{cases} 
   0, & \text{at noise and LF region}, \\  
   \beta\cdot w_n(u, v), & \text{elsewhere.} 
   \end{cases}
       \label{eq:Eq3}
\end{equation} 
   
where $\beta$ is a constant that represents the threshold for controlling the magnitude of the frequency spectrum.

\subsection{Final Formulation of ADFL}

In advanced INR-based methods \cite{Chen_2021_CVPR,Lee_2022_CVPR,Wei_2023_CVPR,Chen_2023_CVPR}, the objective function typically involves only spatial domain loss. 
Yet, solely spatial domain objective functions lead networks to better learn low-frequency components compared to high-frequency ones \cite{32,Isola_2017_CVPR}. Therefore, we propose a frequency domain loss function that adapts to the frequency magnitudes of input images.

Given \(M\) data points from \(N\) images, such as \((x_m, I_n^{HR}(x_m))\), the implicit neural function focuses solely on learnable parameters in the image domain, defined as:

\begin{equation}
\fontsize{9pt}{\baselineskip}\selectfont
  L_{spatial}  = \arg\min_\Theta \sum_{m, n}^{M, N} \Vert {I_n^{HR}(x_m)-s(x_m, I^{LR}_n;\Theta)}\Vert_1.
 \label{eq:Eq4}
\end{equation}

Practically, $X$ range in $[-H,H]$ and $[-W,W]$ for 2D image domain, Note that in \cite{Chen_2021_CVPR,Lee_2022_CVPR}, a decoding function ($f_\theta$) shared by all images with trainable weight $\Theta$.

Above equations and notations illustrates the final loss for frequency loss on the basis of DCT, dealing with distinguishable frequency spectrum by element-wise multiplication $\odot$ with frequency distance matrix $(FDM)$ as Eq.~\ref{eq:Eq2}, reformulate as:

 \begin{equation}
\fontsize{9pt}{\baselineskip}\selectfont
   L_{ADFL}  = \frac {1}{HW} \sum^{H-1}_{u=0}\sum^{W-1}_{v=0}FDM(u, v) \odot M_{DCTL}(u, v).
 \label{eq:Eq5}
\end{equation}  
 
\subsubsection{Problem formulation.}

Eventually, after passing through a decoder, a spatial loss and a global frequency loss defined as the final objective function that could improve widely used in Implicit Neural Function (INF) for SR, where indicates as:

 \begin{equation}
   L_{total}  = L_{spatial}+\lambda \cdot L_{ADFL}
    \label{eq:Eq6}
\end{equation}       

 where $\lambda$ indicated the hyper-parameter balanced the trade-off between the contribution of two losses. 
 
 \subsection{Enhanced Receptive Field Encoder}
For inference, As depicted in Fig.~\ref{fig:Fig6}, expanding the receptive field of encoders helps mitigate frequency inconsistencies caused by aliasing artifacts in low-resolution images and enhances the robustness of INR-based models. 

Inspired by SRFormer \cite{Zhou_2023_ICCV}, we first convert low-resolution RGB images into embeddings using the pixel embedding layer. These embeddings are then refined with a feature encoder that incorporates permuted self-attention blocks (PABs). PABs optimize self-attention by partitioning feature maps into windows and reducing the channel dimension to $C/r^2$, where $C$ is the channel dimension after feature embedding and $r$ is the scale ratio. Spatial tokens are then permuted into the channel dimension. This approach reduces the window size $S$ to $S/r \times S/r$ while preserving the channel dimension, thus effectively balancing computational efficiency and feature richness. Details of this operation and its implementation are provided in the supplementary materials.

 \subsection{Network Detail}

As illustrated in Fig.~\ref{Fig:fig1}, the FreqINR architecture primarily consists of two phases. This section introduces the inference backbone, describes our masked DCT loss, and explains how it reformulates the objective function during the training phase.

\subsubsection{Inference Phase.}
In the inference phase, the key components of our INR-based arbitrary-scale SR network are an encoder $(E_\varphi)$, a local implicit neural representation, a decoder $(f_\theta)$, and an LR skip connection. The implementation follows the methods described in INR-based approaches. For more details, see the supplementary materials.

\subsubsection{Training Phase.}
In the training phase, after obtaining the target HR image, we map target and source image pairs into the relative frequency domain. Although the relative frequency coordinates differ from those in the image domain, they result in a single-channel representation with dimensions $r_xH \times r_yW \times 1$, matching the HR image size. To perform element-wise multiplication for calculating the frequency weighting loss, we create a frequency-adaptive mask with the same dimensions $r_xH \times r_yW \times 1$ as the frequency features. Finally, we combine both spatial loss and frequency loss to form the objective function for training the network.

\section{Experiment}

\subsection{Training}

\subsubsection{Datasets and Evaluation Metrics.}
We use the DIV2K training dataset \cite{agustsson2017ntire} for training. The DIV2K validation set \cite{agustsson2017ntire}, Set5 \cite{bevilacqua2012low}, Set14 \cite{zeyde2012single}, B100 \cite{937655}, and Urban100 \cite{Huang_2015_CVPR} are used for evaluating the performance of our method. We measure model performance using peak signal-to-noise ratio (PSNR) \cite{Chen_2021_CVPR, Lee_2022_CVPR, Wei_2023_CVPR, Chen_2023_CVPR}, and we also crop boundaries to avoid edge effects.

\subsubsection{Implementation Details.}
To ensure fairness, we apply the same implementation details for all baseline INR-based models as described in their respective papers \cite{Chen_2021_CVPR, Lee_2022_CVPR, Wei_2023_CVPR, Chen_2023_CVPR}. For detailed implementation of each backbone model, please refer to the supplementary materials.

\subsection{Evaluation}
In this section, we first evaluate the core technology, Adaptive Discrete Cosine Transform Frequency Loss (ADFL), in terms of effectiveness. Next, we test the FreqINR framework with an enhanced receptive field encoder. Ablation studies assess the impact of ADFL components and the encoder. Finally, we demonstrate ADFL’s versatility by applying it to other tasks.
\begin{figure*}[t]
  \centering
    \belowrulesep=0pt
    \aboverulesep=0pt
     \includegraphics[width=\linewidth]{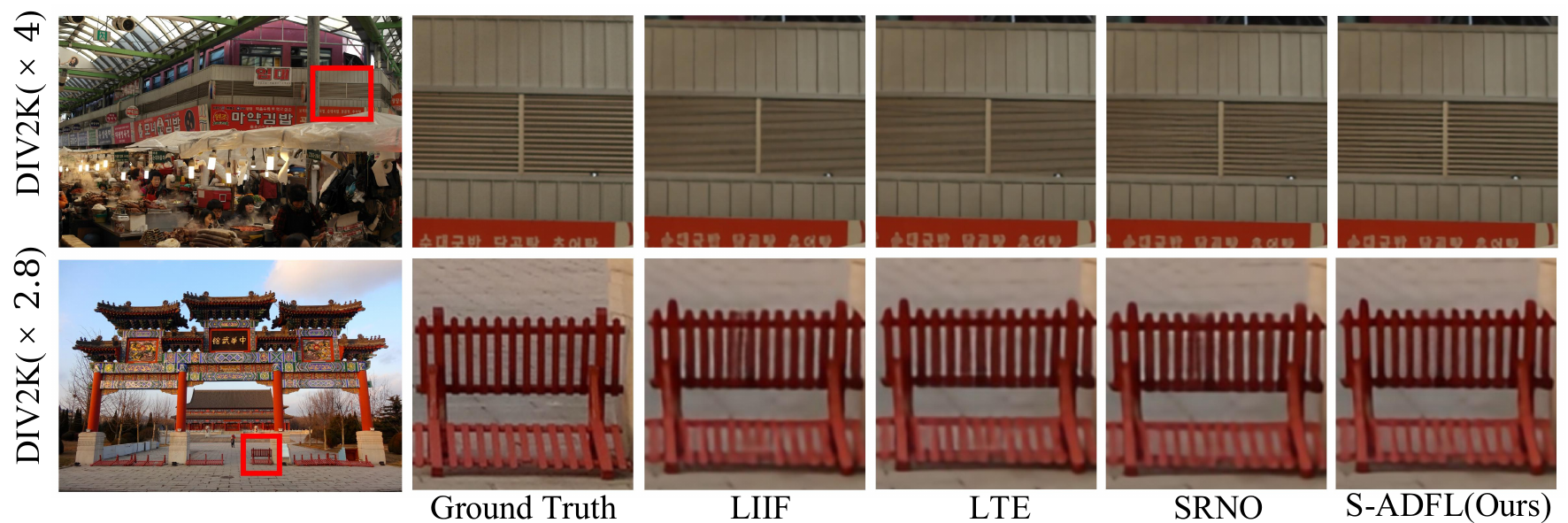}
\caption{Visual comparison of INR-based methods at integer and non-integer scales. The first column highlights close-ups in red. All methods use the EDSR-baseline encoder, trained on DIV2K with random scales ranging from $\times 1$ to $\times 4$.}
    \label{fig:Fig3}
\end{figure*}

\begin{figure*}[t]
  \centering
    \belowrulesep=0pt
    \aboverulesep=0pt
    \includegraphics[width=\linewidth]{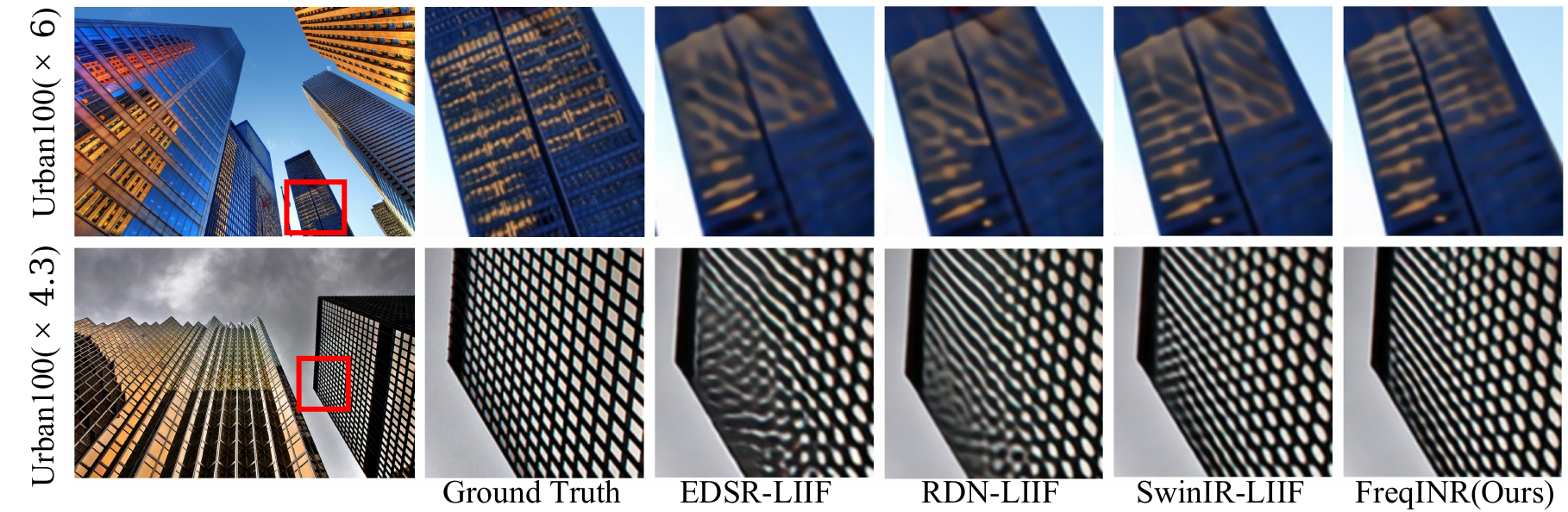}
\caption{Visual comparison of different encoders. All methods use LIIF as the INR-based model during inference.}
      \label{fig:Fig4}
\end{figure*}
\begin{table*}[t]
    \centering
    \small 
    \setlength{\tabcolsep}{12pt} 
    \begin{tabular}{c|ccc|ccccc}
    \toprule
    \multirow{2}{*}{Method}                             & \multicolumn{3}{c|}{In-Scale}       & \multicolumn{5}{c}{Out-of-Scale}  \\
     & \hspace{5pt} ×2 \hspace{5pt} & \hspace{5pt} ×3 \hspace{5pt} & \multicolumn{1}{c|}{\hspace{5pt} ×4 \hspace{5pt}}    & \hspace{5pt} ×6 \hspace{5pt}     & \hspace{5pt} ×12 \hspace{5pt}    & \hspace{5pt} ×18 \hspace{5pt}   & \hspace{5pt} ×24 \hspace{5pt}   & \hspace{5pt} ×30 \hspace{5pt}   \\ 
    \midrule
     EDSR-baseline-LIIF & 34.67 & 30.96 & 29.00 & 26.75  & 23.71  & 22.17 & 21.18 & 20.48 \\
     +ADHL & 34.71 & 31.01 & 29.04 & 26.77  & 23.72  & 22.20 & 21.20 & 20.51 \\
    \midrule
    EDSR-baseline-LTE & 34.72 & 31.02 & 29.04 & 26.81  & 23.78  & 22.23 & 21.24 & 20.53 \\
     +ADHL & 34.77 & 31.07 & 29.09 & 26.81  & 23.76  & 22.23 & 21.23 & 20.53 \\
    \midrule
    EDSR-baseline-CLIT & 34.82 & 31.14 & 29.17 & 26.93  & 23.85  & 22.30 & 21.27 & 20.54 \\
     +ADHL & \underline{34.87} & \underline{31.20} & \underline{29.22} & \underline{26.97}  & \bfseries 23.88  & \bfseries 22.32 & \underline{22.29} & \underline{20.56}\\
    \midrule
    EDSR-baseline-SRNO & 34.85 & 31.11 & 29.16 & 26.90  & 23.84  & 22.29 & 21.27 & 20.56 \\
     +ADHL &\bfseries  {34.91} & \underline {31.20} & \bfseries {29.25} & \bfseries {26.95}  &  \underline{23.85}  & \underline{22.30} & \bfseries {21.32} & \bfseries {20.62} \\
  \bottomrule
\end{tabular}
    \caption{Quantitative comparison on DIV2K validation set (PSNR (dB)). All methods employ EDSR-baseline as the encoder.}
    \label{tab:Tab0}
\end{table*}

\begin{table*}[t]
    \centering
    \small 
    \setlength{\tabcolsep}{12pt} 
    \begin{tabular}{c|ccc|ccccc}
    \toprule
    \multirow{2}{*}{Method}                             & \multicolumn{3}{c|}{In-Scale}       & \multicolumn{5}{c}{Out-of-Scale} \\
     & \hspace{5pt} ×2 \hspace{5pt} & \hspace{5pt} ×3 \hspace{5pt} & \multicolumn{1}{c|}{\hspace{5pt} ×4 \hspace{5pt}}    & \hspace{5pt} ×6 \hspace{5pt}     & \hspace{5pt} ×12 \hspace{5pt}    & \hspace{5pt} ×18 \hspace{5pt}   & \hspace{5pt} ×24 \hspace{5pt}   & \hspace{5pt} ×30 \hspace{5pt}   \\
    \midrule
    Bicubic & 31.01 & 28.22 & 26.66 & 24.82  & 22.27  & 21.00 & 20.19 & 19.59 \\
    EDSR-baseline & 34.55 & 30.90 & 28.94 & -  & -  & - & - & - \\
    EDSR-baseline-MetaSR & 34.64 & 30.93 & 28.92 & 26.61  & 23.55  & 22.03 & 21.06 & 20.37 \\
    EDSR-baseline-LIIF & 34.67 & 30.96 & 29.00 & 26.75  & 23.71  & 22.17 & 21.18 & 20.48 \\
    RDN-LIIF & 34.99 & 31.26 & 29.27 & 26.99  & 23.89  & 22.34 & 21.31 & 20.59 \\
    SwinIR-LIIF & 35.17 & 31.46 & 29.46 & 27.15  & 24.02  & 22.43 & 21.40 & 20.67 \\  
    \bfseries FreqINR-LIIF(Ours)  & \bfseries 35.35 & \bfseries 31.65 & \bfseries 29.63 & \bfseries 27.23  & \bfseries 24.08  & \bfseries 22.49 & \bfseries 21.46 & \bfseries 20.72 \\
  \bottomrule
\end{tabular}
    \caption{Quantitative comparison of LIIF-based methods on DIV2K validation set (PSNR (dB)).}
    \label{tab:Tab1}
\end{table*}

\subsubsection{Quantitative result.}

Tab.~\ref{tab:Tab0} demonstrates the enhancement achieved by integrating Adaptive DCT Frequency Loss (ADFL) into existing INR-based methods \cite{Chen_2021_CVPR, Lee_2022_CVPR, Wei_2023_CVPR, Chen_2023_CVPR}. Remaining the same backbone encoder EDSR-baseline \cite{Lim_2017_CVPR_Workshops}, ADFL significantly improves performance across all prior INR-based methods, particularly for scales of $\times$2, $\times$3, and $\times$4. Regarding the contribution of advanced INR-based methods, ADFL clarifies its potential by narrowing frequency gaps, particularly at in-distribution scales. In Tab.~\ref{tab:Tab1}, to ensure a fair comparison, we utilize LIIF \cite{Chen_2021_CVPR} as the benchmark on the DIV2K validation set. Integrating ADFL with LIIF and extending the encoder’s receptive field, FreqINR delivers state-of-the-art(SOTA) performance across all scaling factors.

The ablation study underscores the advantages of DCT bases and AFWM within ADFL. Experiments with alternative encoders and INR-based methods, tested on Set5 \cite{bevilacqua2012low}, Set14 \cite{zeyde2012single}, B100 \cite{937655}, and Urban100 \cite{Huang_2015_CVPR} list in the supplementary materials.

\subsubsection{Qualitative results.}

Fig.~\ref{fig:Fig3} presents our results at both integer and non-integer scales. Tab.~\ref{tab:Tab0} illustrates that SRNO \cite{Wei_2023_CVPR} performs best. Then, we use SRNO as the baseline and integrate it with ADFL, which is displayed as 'S-ADFL' in the last column. In Fig.~\ref{fig:Fig3}. LIIF \cite{Chen_2021_CVPR} and LTE \cite{Lee_2022_CVPR} focus on dominant frequencies but miss consistent high-frequency details. While SRNO \cite{Wei_2023_CVPR} is effective, it still exhibits texture irregularities. Our method, combining SRNO \cite{Wei_2023_CVPR} with ADFL, closely matches the GT and excels in edge recovery, enhancing high-frequency learning. Fig.~\ref{fig:Fig4} covers building examples from Urban100. Tab.~\ref{tab:Tab1} reveals that FreqINR, with extending RF encoder and our ADHL, effectively reduces spectral discrepancies. Our method is the only one that reconstructs the window shape correctly, while others fail to capture dominant frequencies. FreqINR achieves a uniform texture closely matching the desired effect. Maintaining spectral consistency between the HR and GT images is crucial for precise high-frequency estimation and detail refinement.

In summary, extending the RF of the encoder improves feature capture, and the loss function impacts visual accuracy. The ablation study explores these factors' effects on output images. Additional visual results are in the supplementary materials.

\subsection{Ablation study}

\subsubsection{Effectiveness of DCT Bases and AFWM.}

\begin{figure}[H]
  \centering
      \belowrulesep=0pt
    \aboverulesep=0pt
    \includegraphics[width=\linewidth]{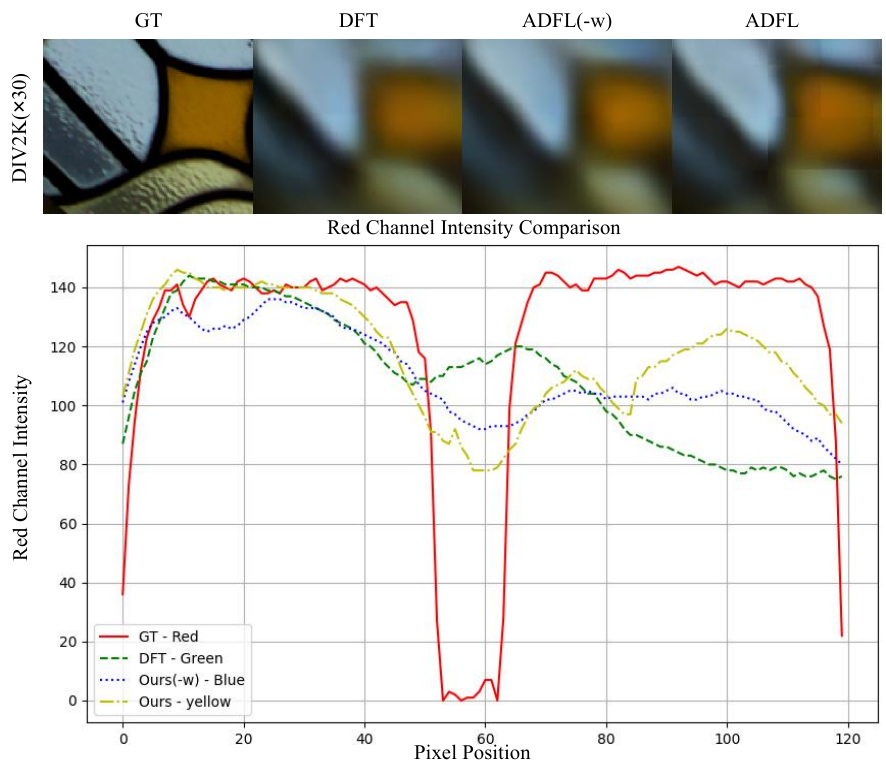}
      \caption{Visual comparison of DFT, DCT, and adaptive DCT at a $\times 30$ scaling factor. The line chart below shows relative pixel intensity variations along the horizontal axis.}
    \label{fig:Fig5}
\end{figure}

Fig.~\ref{fig:Fig5} illustrates that both AFWM and AFWM(-w) without weight matrix improve large-scale SR ($\times 30$) by clearly reconstructing the curves within the orange rectangle. The line chart shows that AFWM(-w) mitigates Gibbs oscillations, and using the weight matrix further aligns the model with the ground truth. Despite the similarities between DCT and DFT losses, our adaptive method excels in recovering fine details for large-scale SR. Additional quantitative results are available in supplementary materials.

\subsubsection{Effectiveness of Extending Receptive Field.}

\begin{figure}[H]
  \centering
    \belowrulesep=0pt
    \aboverulesep=0pt
    \includegraphics[width=\linewidth]{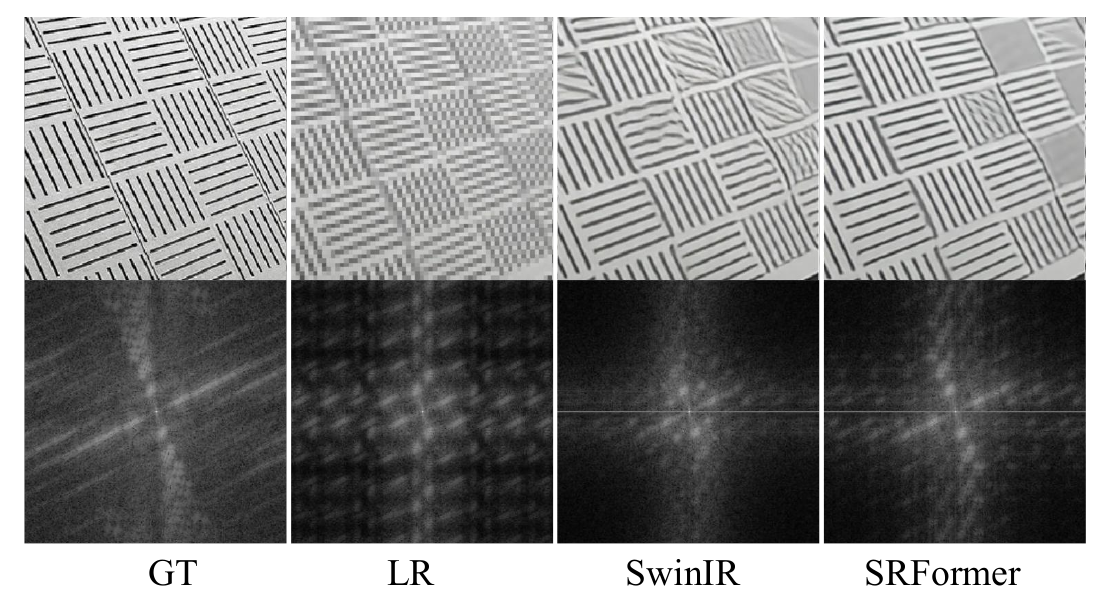}
  \caption{llustration of the influence of encoder's RF.}
      \label{fig:Fig6}
\end{figure}

We discuss the benefits of extending the receptive field (RF) using LIIF as an INR-based model, focusing solely on RF changes without integrating frequency loss. Limited RF results in aliasing with large scaling factors ($\times 24$). Fig.~\ref{fig:Fig6} displays ground-truth and HR images in the first row, with their frequency transforms in the second row. The second column shows LR frequencies deviating from GT, while the third row highlights limited RF miss high-frequency details. With an extended RF \cite{Zhou_2023_ICCV}, as shown in the fourth row, our method effectively removes artifacts and sharpens edges compared to SwinIR \cite{Liang_2021_ICCV}. Additional quantitative results are in the supplementary materials.

\begin{table}[t]
    \setlength{\tabcolsep}{4pt}
    \fontsize{9pt}{\baselineskip}\selectfont
    \belowrulesep=0pt
    \aboverulesep=0pt
    \centering
    \begin{tabular}{c|c|c|c}
    \toprule
    \multirow{1}{*}{Methods}     & \multicolumn{1}{c|}{\#Params.(M)} & \multicolumn{1}{c|}{\#FLOPs(G)}& \multicolumn{1}{c}{\#Mem.(M)} \\
    \midrule
    EDSR-baseline-LIIF & 1.6 & 85.0 & 11.98  \\
        RDN-LIIF & 22.4 & 765.2 & 171.24   \\
        SwinIR-LIIF & 11.9 & 424.6 &113.97\\
        FreqINR-LIIF(Ours) & 10.5 &377.9 &109.92\\
           \midrule
    EDSR-baseline-LTE & 1.7 & 75.3 & 13.13 \\
        RDN-LTE & 22.5 & 755.5 & 172.36   \\
        SwinIR-LTE & 12.1 & 414.8 &115.09 \\
        FreqINR-LTE(Ours) & 10.7 &368.2 &111.04\\
  \bottomrule
\end{tabular}
\caption{Comparison of memory parameters, FLOPs, and memory usage for a ×4 SR task with an input size of $128^2$.}
    \label{tab:Tab2}
\end{table}

\subsubsection{Computation consuming.}

In practical SR applications, fast computation and ample memory are essential, especially for high-quality images like DIV2K. Tab.~\ref{tab:Tab2} compares model demands on an NVIDIA Tesla V100 32GB. We use a window size of 16, while SwinIR \cite{Liang_2021_ICCV} uses 8. Tab.~\ref{tab:Tab1} shows that FreqINR-LIIF outperforms others by expanding the encoder’s receptive field and efficiently transferring spatial information to the channel dimension, as inspired by SRFormer \cite{Zhou_2023_ICCV}. Although frequency loss doesn’t impact computational usage, our extended encoder design offers a more GPU-friendly structure.

\subsection{Discussion}
\subsubsection{Frequency consistency in Arbitrary-Scale SR.}

We proof that adaptively minimizing frequency distance in Arbitrary-Scale Super-Resolution (ASSR) alleviate artifacts and distortion in HR images. Tabs.~\ref{tab:Tab0} and \ref{tab:Tab1} indicate that joint frequency loss improves performance at in-distribution scales. For larger scales, expanding the RF of encoders and refining latent representation are more effective. In future, we aim to unify optimization across scales by combining frequency alignment with multi-scale frequency features.

\subsubsection{DCT in Image Reconstruction.}

Replacing the original FFL \cite{Jiang_2021_ICCV} with our Adaptive DCT Frequency Loss (ADFL) in image reconstruction tasks on DTD \cite{Cimpoi_2014_CVPR} and CelebA \cite{Liu_2015_ICCV} datasets significantly improved PSNR, SSIM, and reduced LPIPS, resulting in more natural images. These results suggest potential for ADFL in tasks like sketch-to-image conversion. Full results on other tasks are provided in the supplement.

\section{Conclusion}

In this paper, we propose FreqINR, a generic framework for arbitrary-scale super-resolution (SR). Our method enhances INR-based approaches by adaptively maintaining frequency consistency across all SR scales during training and inference. During training, we introduce Adaptive DCT Frequency Loss (ADFL) to minimize spectral discrepancies between generated and ground-truth images by dynamically adjusting frequency weights. During inference, expanding the encoder’s receptive field improves model robustness. Extensive experiments show that FreqINR surpasses existing methods in both performance and visual quality while reducing computational costs. We also explore the potential of applying ADFL to other tasks and suggest that advanced optimization techniques could further enhance frequency domain handling across scales.

\bibliography{aaai25}

\end{document}